\documentclass[aps,prl,reprint,amsmath,amssymb,groupedaddress]{revtex4-2}
\usepackage{graphicx}
\usepackage[colorlinks,
linkcolor=blue,
anchorcolor=blue,
citecolor=blue,urlcolor=blue]{hyperref}
\begin{document}

\preprint{APS/123-QED}

\title{The equilibrium distribution function for strongly nonlinear systems}
%\thanks{A footnote to the article title}%

\author{Jialin Zhang}
\author{Yong Zhang}
\author{Hong Zhao}%
\email{zhaoh@xmu.edu.cn}
\affiliation{%
 \\Department of Physics, Xiamen University, Xiamen 361005, Fujian, China \\
}%
\date{\today}

\begin{abstract}

The equilibrium distribution function determines macroscopic observables in statistical physics. While conventional methods correct equilibrium distributions in weakly nonlinear or near-integrable systems, they fail in strongly nonlinear regimes. We develop a framework to get the equilibrium distributions and dispersion relations in strongly nonlinear many-body systems, incorporating corrections beyond the random phase approximation and capturing intrinsic nonlinear effects. The theory is verified on the nonlinear Schr\"{o}dinger equation, the Majda–McLaughlin–Tabak model, and the FPUT-$\beta$ model, demonstrating its accuracy across distinct types of nonlinear systems. Numerical results show substantial improvements over existing approaches, even in strong nonlinear regimes. This work establishes a theoretical foundation for equilibrium statistical properties in strongly nonlinear systems.

\end{abstract}
\maketitle

\textbf{\emph{Introduction.——}}
Statistical physics provides the foundation for computing macroscopic observables. Once the equilibrium distribution is known, ensemble-averaged quantities can be systematically obtained. In practice, these distributions are often those derived near integrable limits, as commonly presented in textbooks. However, nonlinear interactions are ubiquitous in realistic systems. While they promote thermalization\textsuperscript{\cite{onorato2015route,wang2020wave,onorato2023wave,lin2025universality}}, they also generate deviations from near-integrable behavior. In weakly nonlinear regimes, phonon Green's function methods\textsuperscript{\cite{werthamer1970self,tripathi1974self,tadano2015self,tadano2022first,xiao2023anharmonic}} and tree-level approximation\textsuperscript{\cite{boulware1968tree,driesse2024conservative,mougiakakos2024schwarzschild}} rooted in quantum field theory offer perturbative corrections Feynman diagrams. Although high-order corrections are possible in principle, their complexity limits practical use to low-order expansions.

As a representative example, the nonlinear Schr\"{o}dinger equation (NSE) conserves both energy and particle number. Consequently, in the vanishing nonlinearity limit, the mode occupation follows the Rayleigh–Jeans (RJ) distribution:
\begin{equation}
n_k=\frac{k_B T}{\omega_k-\mu} ,
\label{RJwr}
\end{equation}
Here, $\omega_k$ is the mode frequency, $n_k$ is the average occupation number, $T$ is the temperature, $k_B$ is the Boltzmann constant, and $\mu$ is the chemical potential.

For weak nonlinearity, several mean-field-based approaches have been developed to provide effective corrections under specific conditions. These include the variational approach\textsuperscript{\cite{liu2015renormalized,liu2016variational}}, the self-consistent harmonic approximation\textsuperscript{\cite{koehler1966theory,bruesch2012phonons}}, the self-consistent phonon theory\textsuperscript{\cite{he2008thermal,he2009origin,he2016quantum,masuki2022anharmonic}}, and trivial-resonance renormalization methods from wave turbulence theory\textsuperscript{\cite{nazarenko2011wave,gershgorin2005renormalized,leisman2019effective,lee2009renormalized,gershgorin2007interactions,lvov2018double,chibbaro20184}}. These methods correct physical quantities by modifying dispersion relations or phonon frequencies from a mean-field perspective. They are based on a common renormalization idea: nonlinear interactions are assumed not to alter the system’s structure. A partially integrable component is extracted from the non-integrable Hamiltonian to construct an effective Hamiltonian, leaving the remaining interactions as small perturbations. This effectively reduces strong interactions to weak ones, enabling treatment within perturbative frameworks.

Among these approaches, wave turbulence theory yields an analytic expression for the frequency shift $\tilde{\omega}_k$ arising from trivial resonances. This shift modifies the dispersion relation and leads to a correction of the RJ distribution:
\begin{equation}
n_k=\frac{k_B T}{\tilde\omega_k-\mu},
\label{RJtr}
\end{equation}

Conventional approaches, including wave turbulence theory, typically invoke the random phase approximation (RPA), often supplemented by the random amplitude assumption—an analog of the molecular chaos hypothesis. Within this framework, nonlinear interactions are treated as white noise acting on each mode. We show that, under RPA, the averaged contribution of residual nonlinear interactions to the frequency shift vanishes. Thus, the trivial-resonance correction emerges as the maximal correction permissible within the RPA framework. However, these methods are intrinsically limited to near-integrable regimes—i.e., systems that can be effectively rendered weakly nonlinear via renormalization or coordinate transformations.

For strongly or intrinsically nonlinear systems—such as those with odd-order interaction potentials—the RPA fails, as we demonstrate in this work. In these regimes, Wick's theorem  no longer apply. We develop an analytical framework to determine the equilibrium distribution function beyond the RPA, capturing corrections of colored noise. Starting from a generalized equipartition principle, we derive the equilibrium statistics for a broader class of nonlinear systems:
\begin{equation}
n_k=\frac{k_B T}{\tilde\omega_k-\mu-\omega^*_k},
\label{RJre}
\end{equation}
where, $\omega^*_k$ is the correction arising from interactions beyond the RPA. It does not correspond to a uniform shift of individual mode frequencies. Instead, it reflects averaged modifications to the spectral structure, including features such as frequency splitting arising in systems with odd-order nonlinearities. Consequently, this correction allows for qualitative changes in the spectrum and captures the essential impact of intrinsic nonlinear interactions, marking a clear departure from near-integrable behavior.

To illustrate the generality of our approach, we also derive correction formulas for both the Majda–McLaughlin–Tabak (MMT) model\textsuperscript{\cite{majda1997one}} and the FPUT-$\beta$ equation\textsuperscript{\cite{fermi1955studies}}. The NSE and the MMT model are complex-valued systems, with the latter characterized by a more general dispersion relation. In contrast, the FPUT-$\beta$ model represents a real-valued system. Numerical simulations confirm that our theoretical predictions remain accurate even in strongly nonlinear regimes, substantially outperforming conventional approaches.

\textbf{\emph{NSE.——}} The generalized equipartition theorem imposes strong constraints on nonlinear interactions: at equilibrium, the distribution must satisfy
\begin{equation}
    \begin{aligned}
     \langle x_i \frac{\partial H-\mu N}{\partial x_j} \rangle =k_B T \delta_{ij}
\end{aligned}
\label{GE}
\end{equation}
where, $x_i$ is the generalized coordinate.

As a representative case, we apply the generalized equipartition theorem to the discrete NSE with quartic nonlinearity, deriving its equilibrium distribution and the corresponding averaged frequency.The Hamiltonian is given by:
\begin{equation}
    \begin{aligned}
    H &= \sum_{l\in \Lambda_L}{|\psi_{l+1}- \psi_l|^2 +\frac{b}{2} |\psi_l|^4},\\
    \end{aligned}
    \label{HNLSE}
\end{equation}
where, $\psi_l \in \mathbb{C}$, and the spatial domain is $\Lambda_L = \{1, 2, \dots, L\}$. At equilibrium, we expand $\psi_l$ in the eigenbasis $a_k$ ($k = 1, 2, \dots, L$) of the linear term in Eq.~\ref{HNLSE}, assuming periodic boundary conditions. This yields the Hamiltonian in the form
\begin{equation}
    \begin{aligned}
    H &=\sum_k {\omega_k a_k a^*_k} + \lambda\sum_{1234} {a_1 a_2 a^*_3 a^*_4 \delta^{12}_{34}}\\
    \end{aligned}
\end{equation}
where $\delta^{12}_{34} \equiv \delta^{k_1 + k_2}_{k_3 + k_4}$, $\lambda = b/2L$, and $\omega_k = 4 \sin^2\left( \frac{k\pi}{L} \right)$ is the dispersion relation.

The system conserves both the Hamiltonian $H$ and the total particle number $N = \sum_l |\psi_l|^2$. therefore its partition function is $Z=\int e^{-\beta (H-\mu N)}d\Omega$, where $d\Omega=dx_1dx_2\cdots$, $\beta=\frac{1}{k_B T}$. Treating $(a_k, a_k^*)$ as generalized coordinates and considering only the linear part of the Hamiltonian, the generalized equipartition theorem yields
\begin{equation}
    \begin{aligned}
     n_k (\omega_k - \mu) =k_B T
    \end{aligned}
    \label{RJnonrenormalized}
\end{equation}
where $n_k = \langle a_k a_k^* \rangle$. This corresponds to Eq.~\ref{RJwr}.

\begin{figure}[]
    \centering
    \includegraphics[scale=0.5]{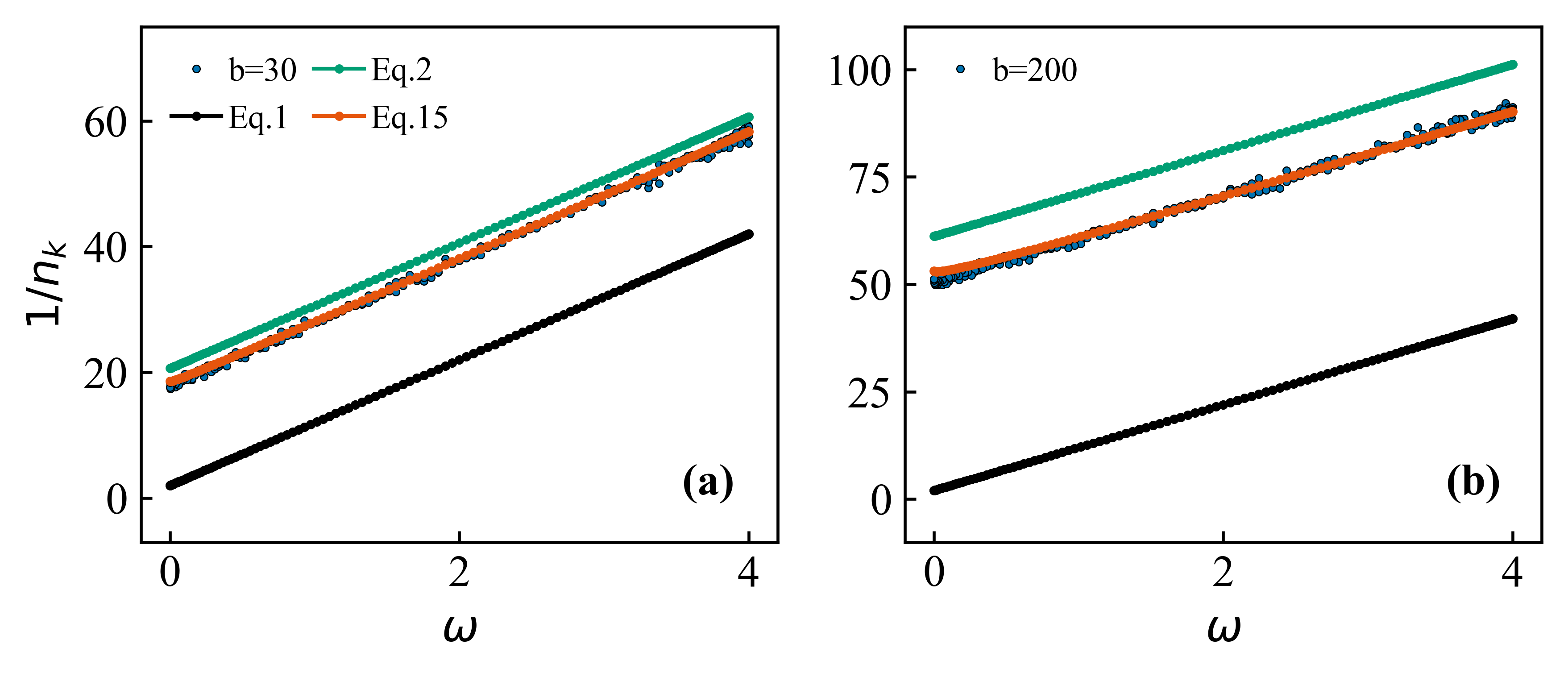}
    \caption{
    For the NSE with $b = 30$(a) and $b =200$(b), Comparison between simulated distributions (dots) and  the linear prediction (Eq.\ref{RJwr}, black), the trivial resonance prediction  (Eq.\ref{RJtr}, green), and our prediction (Eq.~\ref{RJNLSE}, red). 
    }
    \label{NLSERJFIG}
\end{figure}

For demonstration purposes, we perform numerical simulations of the NSE model with $T = 0.1$, $\mu = -0.2$,  $L = 256$, and set $k_B=1$. In Fig.\ref{NLSERJFIG}, we present the numerically obtained equilibrium distributions for different nonlinear strengths, along with the theoretical prediction from Eq.\ref{RJnonrenormalized}. Significant discrepancies are observed, which increase with the strength of nonlinearity. This discrepancy clearly arises from nonlinear effects—specifically, the quadratic contribution in the nonlinear term has not been extracted.
If we retain only the trivial resonances, i.e., those satisfying $(k_1 = k_3, k_2 = k_4)$ and $(k_1 = k_4, k_2 = k_3)$, the generalized equipartition theorem yields 
\begin{equation}
    \begin{aligned}
    n_k (\tilde{\omega}_k-\mu) +b/L\sum^*_{123} \langle a_1 a_2^* a_3^* a_k\rangle\delta_{12}^{3k}=k_BT
    \end{aligned}
    \label{RJ}
\end{equation}
with $\tilde{\omega}_k = \omega_k + 2bN/L$.
In the weakly nonlinear regime, system variables can be treated as approximately independent and quasi-Gaussian. Under this assumption, the RPA holds, and Wick's theorem applies: $\langle a_1 a_2 a_3^* a_4^* \rangle = n_1 n_2 \delta_{13} \delta_{24}$. This implies that nontrivial resonant interactions contribute negligibly, i.e., $\langle a_1 a_2 a_3^* a_4^* \rangle = 0$. As a result, the RJ distribution is corrected to Eq.~\ref{RJtr}, consistent with predictions from wave turbulence theory and renormalization approaches\textsuperscript{\cite{lee2009renormalized},\cite{nazarenko2011wave}}.

As shown in Fig.~\ref{NLSERJFIG}, the correction from trivial resonances significantly improves the accuracy of the theoretical prediction. Further numerical validation indicates that this correction remains accurate for $b \lesssim 10$. The effect of trivial resonances has a clear physical interpretation: it induces a frequency shift. This corresponds to a renormalization of the integrable Hamiltonian, meaning that one can replace the original Hamiltonian with a new one expressed in terms of shifted-mode variables. With the dispersion relation modified by the frequency correction, the formulas derived from the linear Hamiltonian remain applicable.

\textbf{\emph{RJ distribution corrected by nontrivial resonances.}}
Fig.~\ref{NLSERJFIG} also shows that, in the strongly nonlinear regime, significant discrepancies persist between the Eq.\ref{RJtr} and numerical results. To incorporate the corrections induced by nontrivial resonances, we proceed by renormalizing the Hamiltonian as follows:
\begin{equation}
    \begin{aligned}
    H &=\sum_k {\tilde{\omega}_k a_k a^*_k} + \lambda\sum_{1234}^* {a_1 a_2 a^*_3 a^*_4 \delta^{12}_{34}}\\
    \end{aligned}
    \label{HDNLSk}
\end{equation}
where, $\sum^*$ denotes a summation excluding the trivial resonances. In this regime, the assumption of independent quasi-Gaussian statistics—or equivalently, the random phase approximation—is no longer strictly valid, and the summation term in Eq.\ref{RJ} no longer vanishes.

To compute the contribution from nonlinear interactions not captured by the trivial resonance correction, we introduce two sets of external sources, $\{J^1_k,J^2_k\}\in R$, and couple the system Hamiltonian $H$ to $\sum_k (J^1_k a_k + J^2_k a^*_k)$. This allows us to construct a source-dependent partition function (generating functional):
\begin{equation}
    \begin{aligned}
    Z(J) =& \int e^{-\beta (H -\mu N)+ \sum_k (J^1_k a_k + J^2_k a^*_k)}d\Omega\\
    \end{aligned}
    \label{PartitionFunction}
\end{equation}
with $d\Omega = \prod_k da_k$. Setting $J^1_k = J^2_k = 0$ recovers the original system. From this generating functional, we derive the general form of the fourth-order moment:
\begin{equation}
    \begin{aligned}
    \langle a_1 a_2 a_3^* a_4^* \rangle
    =\frac{1}{Z(0)} \frac{\partial^4 Z(J)}{\partial J^1_1 \partial J^1_2 \partial J^2_3 \partial J^2_4 }\Big|_{J^1_k=J^2_k=0}\\
    \end{aligned}
\end{equation}
By expanding the nonlinear part of the Hamiltonian in Eq.~\ref{PartitionFunction} from the exponent, we obtain (see Supplemental Material for a detailed derivation):
\begin{equation}
    \begin{aligned}
    Z(J) & =\prod_{k} \frac{\pi}{\beta(\tilde{\omega}_k-\mu)}e^{\frac{J^1_k J^2_k}{\beta(\tilde{\omega}_k-\mu)}} -\beta \lambda \sum_{1234}^* \delta^{12}_{34} J^2_1 J^2_2 J^1_3 J^1_4 \\ & \prod_{k \in \{1234\}}\frac{1}{\beta(\tilde{\omega}_k-\mu)} \prod_{k} \frac{\pi}{\beta(\tilde{\omega}_k-\mu)}e^{\frac{J^1_k J^2_k}{\beta(\tilde{\omega}_k-\mu)}}
\end{aligned}
\label{PartitionResult}
\end{equation}
The first term corresponds to the partition function with trivial-resonance corrections, where $Z_0(0)=\prod_{k} \frac{\pi}{\beta(\tilde{\omega}_k-\mu)} \propto \prod_{k}n_k^0$ and $n_k^0=\frac{1}{\beta(\tilde{\omega}_k-\mu)}$. The second term, $Z_1$, accounts for contributions from nontrivial interactions. Since $\langle a_1 a_2 a_3^* a_4^* \rangle_0 =\frac{1}{Z(0)} \frac{\partial^4 Z_0(J)}{\partial J^1_1 \partial J^1_2 \partial J^2_3 \partial J^2_4 }\Big|_{J^1_k=J^2_k=0} =0$, we compute the first-order correction to the partition function:
\begin{equation}
\begin{aligned}
    \langle a_1 a_2 a_3^* a_4^* \rangle_1 &=\frac{1}{Z(0)} \frac{\partial^4 Z_1(J)}{\partial J^1_1 \partial J^1_2 \partial J^2_3 \partial J^2_4 }\Big|_{J^1_k=J^2_k=0} \\
        &\simeq  -4\beta \lambda \delta^{12}_{34} \prod_{k \in \{1234\}}\frac{1}{\beta(\tilde{\omega}_k-\mu)} \\
    \end{aligned}
    \label{4moment}
\end{equation}
This expression indicates that, for the fourth-order moment, the first-order correction accounts for interactions between two groups of four-wave modes. Furthermore, it becomes evident that when higher-order corrections are considered, strongly nonlinear systems violate Wick’s theorem and no longer consist of independent Gaussian modes. Instead, statistical correlations emerge among four interacting modes, effectively producing colored-noise-like behavior (see Supplemental Material).

In this case, the generalized equipartition theorem yields
\begin{equation}
    \begin{aligned}
    (\tilde{\omega}_k-8\beta \lambda^2 \sum_{123}^* n_1^0 n_2^0 n_3^0\delta^{12}_{3k}  -\mu)n_k \simeq k_B T\\
    \end{aligned}
    \label{GEE-}
\end{equation}
Here, we replace $n_k^0$ in Eq.\ref{GEE-} with $n_k$, introducing only a higher-order error (see SM). This allows us to express $n_k$ as
\begin{equation}
    \begin{aligned}
    n_k = \frac{k_B T}{\tilde{\omega}_k-8\beta \lambda^2 \sum_{123}^* n_1^0 n_2^0 n_3^0\delta^{12}_{3k}  -\mu}
    \end{aligned}
    \label{RJNLSE}
\end{equation}
which corresponds to Eq.\ref{RJre}, where all quantities on the right-hand side are analytically computable. As shown in Fig.~\ref{NLSERJFIG}, this correction significantly improves the accuracy of theoretical predictions, with excellent agreement with numerical results up to $b \gtrsim 200$.

\begin{figure}[]
    \centering
    \includegraphics[scale=0.5]{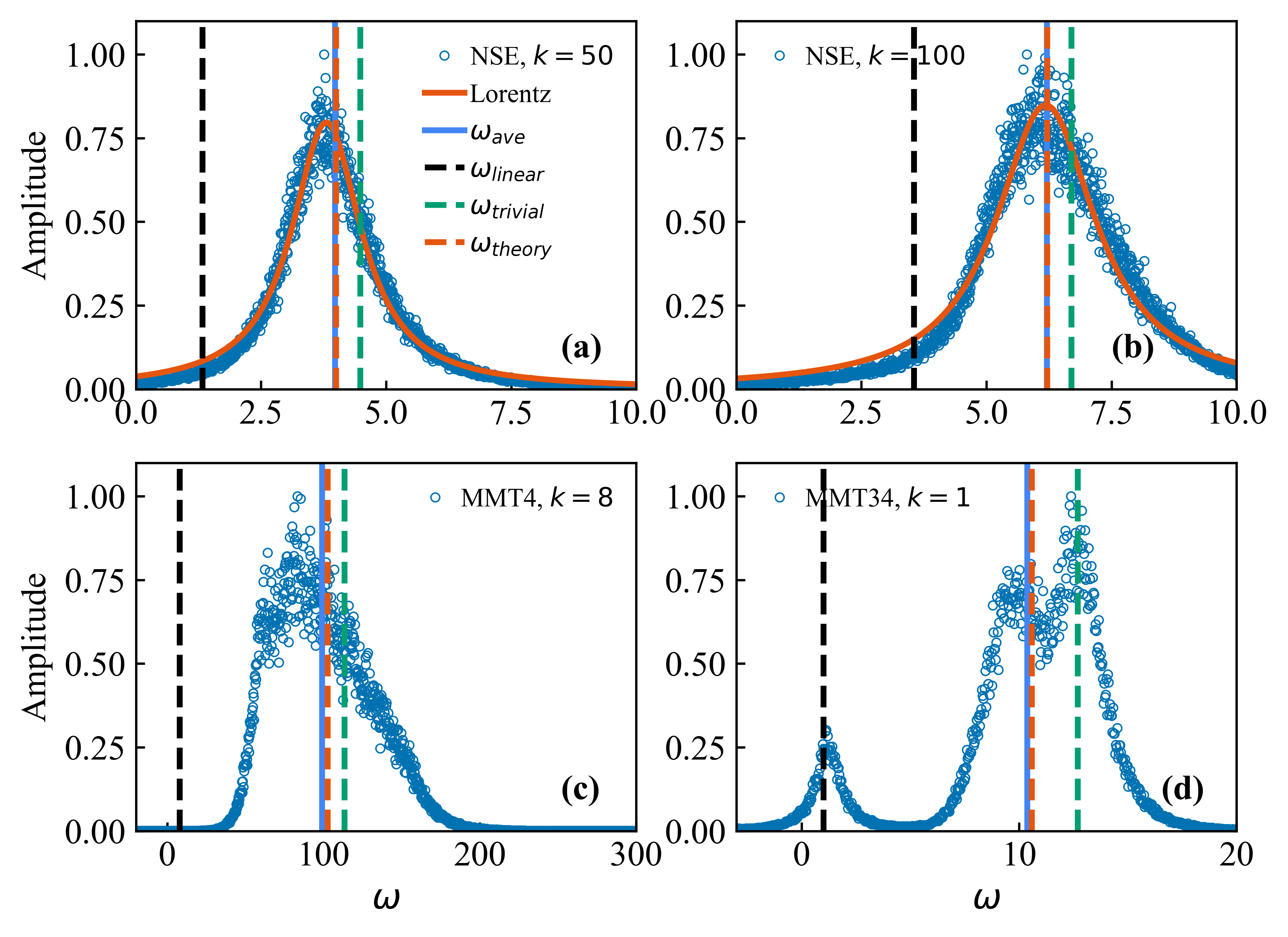}
    \caption{For the NSE at $b = 70$ for $k = 50$(a) and $k = 100$(b); MMT4 model at $T = 10$, $k = 8$ (c); MMT34 model at $T = 10$, $k = 1$ (d): Spectrum (dots), average frequency (blue solid line), unrenormalized linear frequency (black dashed line), trivial-resonance frequency prediction (green dashed line), and our frequency prediction (red dashed line). Spectra are normalized by their peak values. Red solid line in (a) and (b) show Lorentzian fits.}
    \label{frequency}
\end{figure}

\textbf{Average frequency derived from the equilibrium distribution.} For the $k$th mode, the average frequency is defined as$\bar{\omega}_{k}=-\frac{\int \omega S_k(\omega) d\omega}{\int  S_k(\omega) d\omega}$, where $S_k(\omega)$ denotes the power spectral density of $a_k(t)$. From this, one can derive $\bar{\omega}_{k}=\frac{k_B T}{n_k}+\mu$ (see SM). Accordingly, we obtain the frequency correction induced by nontrivial resonances:
\begin{equation}
    \begin{aligned}
    \bar{\omega}_{k} =\tilde{\omega}_k-8\beta \lambda^2 \sum_{123}^* n_1^0 n_2^0 n_3^0\delta^{12}_{3k}
    \end{aligned}
    \label{omegaave}
\end{equation}
Fig.~\ref{frequency} shows that the trivial-resonance correction tends to overestimate the frequency shift, while the nontrivial-resonance contribution corrects the overestimate, yielding a theoretical prediction in excellent agreement with the numerical results.

\textbf{\emph{MMT system.}}
Based on the generalized equipartition theorem, we derive the equilibrium distributions for both the quartic potential MMT model (MMT4)
\begin{equation}
    \begin{aligned}
        &H=\sum_k \omega_k |a_k|^2  + \frac{1}{2} \sum_{1234} I_{1234} a_{k_1} a_{k_2} a^*_{k_3} a^*_{k_4} \delta^{12}_{3k}
    \end{aligned}
\end{equation}
and mixed potential MMT model (MMT34)
\begin{equation}
    \begin{aligned}
        H=&\sum_k \omega_k a_k a^*_k + a \sum_{123} a_1 a_2 a^*_3 \delta^{12}_{3}  + a \sum_{123} a_1 a^*_2 a^*_3\delta^{1}_{23} \\
        & + \frac{b}{2} \sum_{1234}  a_1 a_2 a^*_3 a^*_4 \delta^{12}_{34} \\
    \end{aligned}
\end{equation}

where $\omega_k = |k|^m$ is the dispersion relation. In the MMT4 model, $I_{1234} = |k_1 k_2 k_3 k_4|^{n/4}$ characterizes the strength of the nonlinear interaction. In the MMT34 model, the coefficient $a$ controls the cubic term, while $b/2$ characterizes the quartic term. In the canonical ensemble, the generalized equipartition theorem yields, in the vanishing nonlinearity limit, the distribution function $\langle |a_k|^2 \rangle = n_k = \frac{k_B T}{\omega_k}$. When only the trivial-resonance correction from the quartic term is included, this becomes $n_k = \frac{k_B T}{\tilde{\omega}_k}$, where $\tilde{\omega}_k = |k|^m + \left( 2 \sum_{k'} |k'|^{n/2} \langle |a_{k'}|^2 \rangle \right) |k|^{n/2}$.

For the MMT4 model, considering nontrivial resonances yields:
\begin{equation}
    \begin{aligned}
        n_k \simeq \frac{k_B T}{\tilde{\omega}_k-2\beta \sum^*_{123} I_{123k}^2 \delta^{12}_{3k} \prod_{k \in \{123\}}\frac{1}{\beta\tilde{\omega}_k} }
    \end{aligned}
    \label{RJMMT4}
\end{equation}

\begin{figure}[]
    \centering
    \includegraphics[scale=0.5]{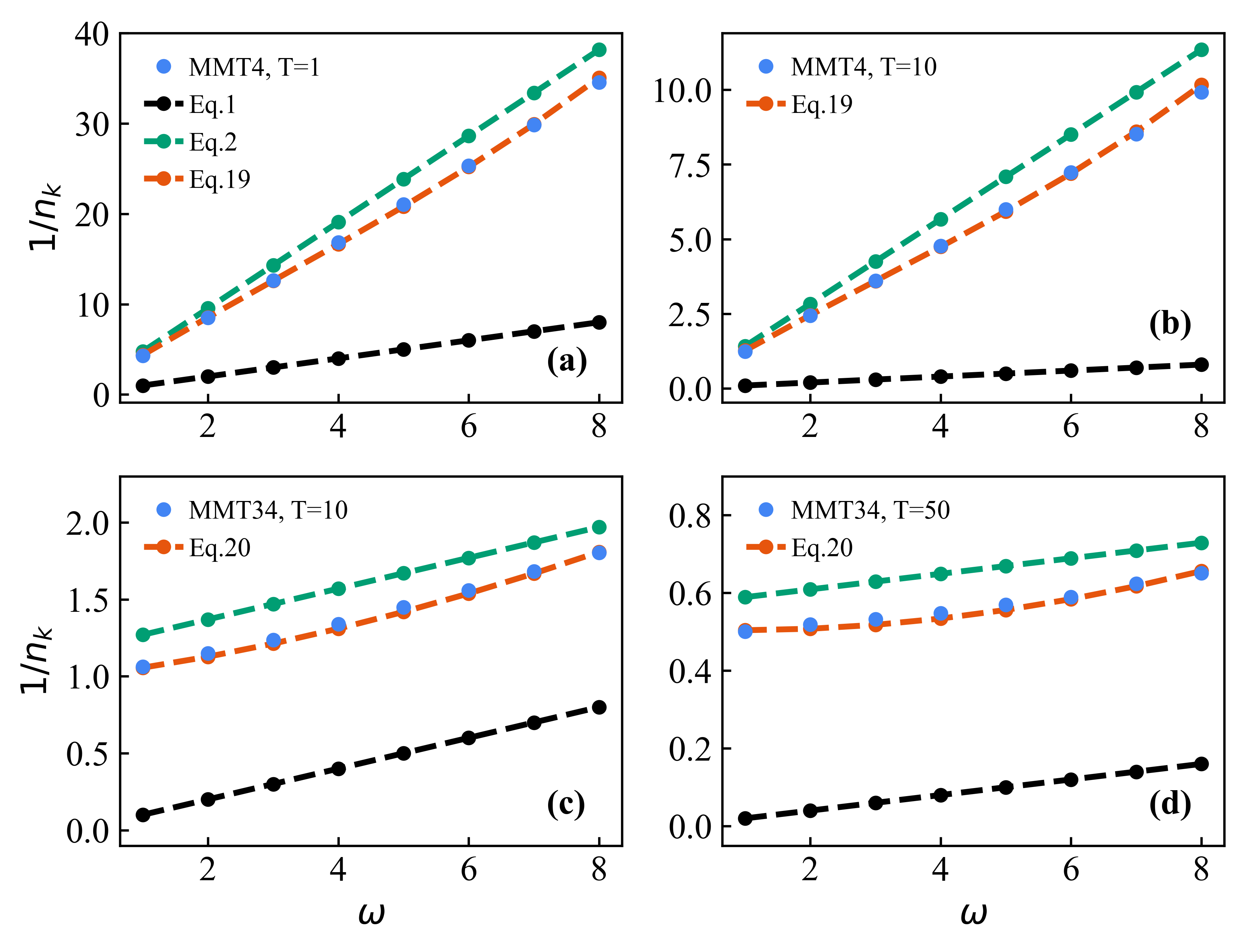}
    \caption{For the MMT4 model with $T=1$(a), $T=10$(b), and MMT34 model with $T=10$(c), $T=50$(d). 
    Comparison between the simulated distribution(dots),the linear prediction (black), the trivial resonance
    prediction (green), {the theoretical prediction from Eq.\ref{RJMMT4} (or Fig.\ref{RJMMT34}) (red)}.}
    \label{MMT4+34}
\end{figure}

For the MMT34 model, the corresponding expression is:
\begin{equation}
    \begin{aligned}
     n_k \simeq  \frac{k_B T}{\tilde{\omega}_k - \beta F_{123}},
    \end{aligned}
    \label{RJMMT34}
\end{equation}
where $F_{123} = \sum_{12} 2 a^2 n_1^0 n_2^0 \delta^{12}_k + \sum_{12} 4 a^2 n_1^0 n_2^0 \delta^{1}_{2k} + \sum_{123}^* 2 b^2 n_1^0 n_2^0 n_3^0 \delta^{12}_{3k}$.
(see SM) For the MMT models, we perform numerical simulations with $m = 1$, $n = 2$, $a = 1$, $b = 1$, $k_{\text{max}} = 8$.
Fig.~\ref{MMT4+34} compares various theoretical predictions with numerical simulations. Even in the strongly nonlinear regime, considering nontrivial resonances yields highly accurate results, significantly outperforming the trivial-resonance correction.

Similar to the NSE, the above formulation also provides a theoretical prediction for the mean value of the frequency. In Fig.\ref{frequency}(c) and (d), we compare the predicted mean values of corresponding frequencies with those obtained from simulations for the MMT4 and MMT34 models, respectively. The agreement between theory and simulation is remarkably good. It is important to emphasize that the correction to the mean frequency is not equivalent to a correction to the spectral peak. As seen in Fig.\ref{frequency}, the mean and peak frequencies do not coincide due to the asymmetry of the spectrum, which deviates from a strict Lorentzian shape. The trivial-resonance correction corresponds to a simple frequency shift, which in the strongly nonlinear regime deviates significantly from both the spectral peak and the mean. Notably, Fig.~\ref{frequency}(d) shows that the frequency of the $k$th mode undergoes splitting due to nontrivial resonances, resulting in three distinct peaks. These peaks are located near the linear eigenfrequency, the trivial-resonance-corrected frequency, and the mean frequency predicted with nontrivial-resonance corrections. This clearly demonstrates that the nontrivial correction cannot be interpreted as a simple frequency shift, but instead reflects a genuinely nonlinear effect.

\begin{figure}[]
    \centering
    \includegraphics[scale=0.5]{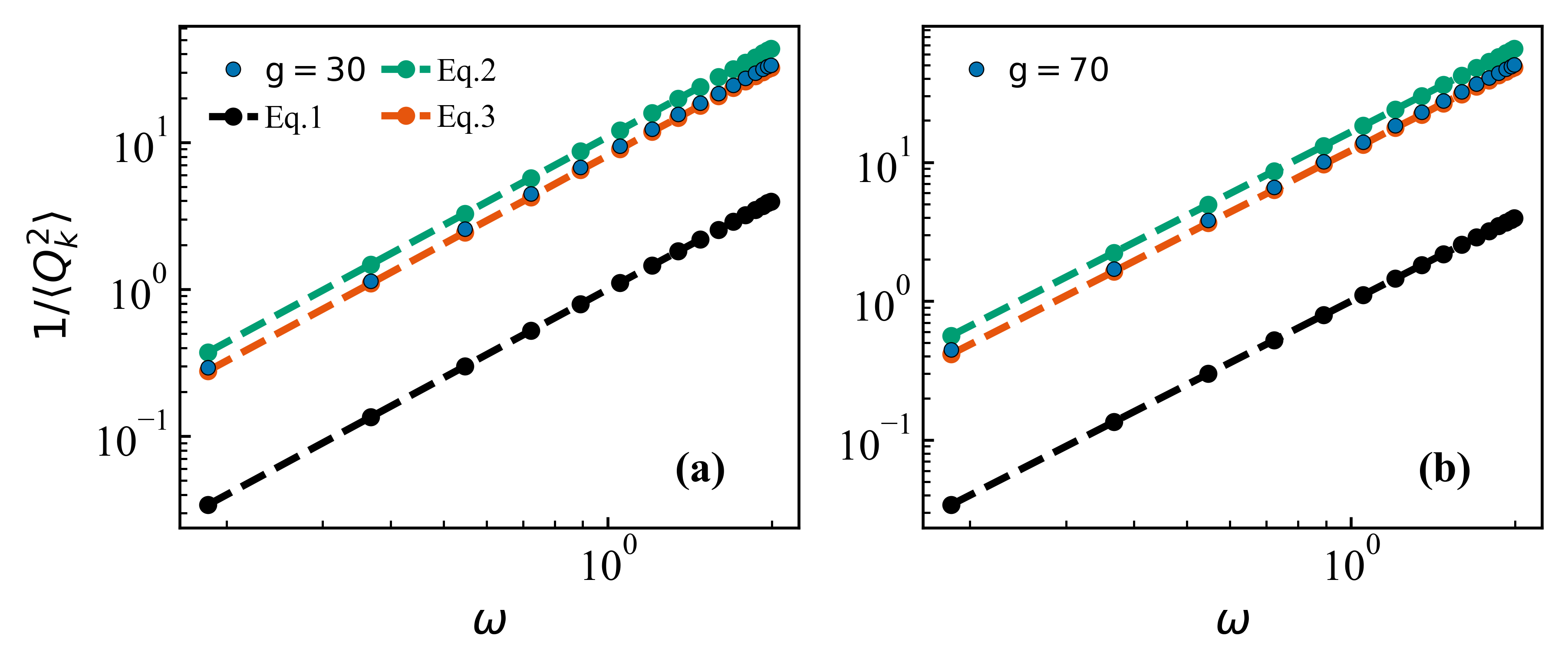}
    \caption{FPUT-$\beta$ system at $b = 30$(a) and $b = 70$(b): Comparison between the simulated distribution(dots),the linear prediction (black), the trivial resonance
    prediction (green), the theoretical prediction including nontrivial resonances (red)}
    \label{FPUT}
\end{figure}

\textbf{\emph{FPUT-$\beta$ system.}}
The Hamiltonian of the FPUT-$\beta$ system\textsuperscript{\cite{fermi1955studies}} is given by
\begin{equation}
    \begin{aligned}
        &H = \sum_j \frac{p_j^2}{2} + \frac{(q_{j+1} - q_j)^2}{2} +  \frac{g(q_{j+1} - q_j)^4}{4}\\
    \end{aligned}
\end{equation}
We transform the Hamiltonian into canonical coordinates and further expand it in the basis corrected by trivial resonances. Based on the generalized equipartition theorem, the correction due to nontrivial resonances can then be derived. For the FPUT-$\beta$ system, we perform numerical simulations with $T=1$, $j_{max}=16$. Fig.\ref{FPUT} shows the results under fixed boundary conditions, where the theoretical predictions remain consistent with numerical simulations even in the strongly nonlinear regime. However, theoretical analysis for this model is more involved, as it requires accounting for multiple types of nontrivial resonances. Detailed derivations are provided in the Supplemental Material.

\textbf{\emph{Conclusion and discussion.}}
In conclusion, we have proposed a framework for correcting the equilibrium distribution of nonlinear systems in regimes where the random phase approximation breaks down and weakly nonlinear assumptions no longer apply. Based on this approach, we derived the corrected distributions for several representative models, including the NSE, the MMT model, and the FPUT-$\beta$ system. Numerical simulations show that the range of nonlinear strength over which our correction remains accurate far exceeds that of the trivial-resonance correction.
Furthermore, using the corrected distributions, we obtain theoretical predictions for the mean values of mode frequencies that agree remarkably well with simulations. These results effectively modify the dispersion relation at equilibrium. This correction goes beyond a simple frequency shift, as it captures the mean values of the frequency modes shaped by intrinsically asymmetric spectral features.

As clearly seen in the case of the MMT34 model, a single frequency can turn into the multi-peaked distribution under strong nonlinearity. This indicates that the number of effective modes in the system no longer corresponds to the canonical linear modes, and the system can no longer be reduced to a structurally equivalent weakly nonlinear one via renormalization. Nevertheless, the corrected equilibrium distribution remains consistent with numerical simulations, demonstrating that physical observables can still be computed as ensemble averages.

% \nocite{*} % 这会引用所有文献

\bibliographystyle{unsrt} % 指定参考文献样式
\bibliography{reference}% Produces the bibliography via BibTeX.

%%%%%%%%%%%%%%%%%%%%%%%%%%%%%%%%%%%%%%%%%%%%%%%%%%%%%%%%%%%%%%%%%%%%%%%%

\end{document}